\documentclass[journal]{IEEEtran}
\ifCLASSINFOpdf
\else
\fi
\usepackage{graphicx}
\usepackage{float}
\usepackage{amssymb}
\usepackage{amsmath}
\usepackage{cite}
\usepackage{multirow}
\usepackage{stfloats}
\begin{document}
\title{Self-supervised Change Detection in Multi-view Remote Sensing Images}
%
\author{Yuxing~Chen,~\IEEEmembership{~}
	    Lorenzo~Bruzzone,~\IEEEmembership{Fellow,~IEEE}
        
\thanks{Y. Chen and L. Bruzzone are with the Department of Information Engineering and Computer Science, University of Trento, 38122 Trento, Italy (e-mail:chenyuxing16@mails.ucas.ac.cn;lorenzo.bruzzone@unitn.it).}
\thanks{Corresponding author: L. Bruzzone}}

\maketitle

\begin{abstract}
The vast amount of unlabeled multi-temporal and multi-sensor remote sensing data acquired by the many Earth Observation satellites present a challenge for change detection.
Recently, many generative model-based methods have been proposed for remote sensing image change detection on such unlabeled data. 
However, the high diversities in the learned features weaken the discrimination of the relevant change indicators in unsupervised change detection tasks.
Moreover, these methods lack research on massive archived images.
In this work, a self-supervised change detection approach based on an unlabeled multi-view setting is proposed to overcome this limitation.
%
This is achieved by the use of a multi-view contrastive loss and an implicit contrastive strategy in the feature alignment between multi-view images.
In this approach, a pseudo-Siamese network is trained to regress the output between its two branches pre-trained in a contrastive way on a large dataset of multi-temporal homogeneous or heterogeneous image patches.
Finally, the feature distance between the outputs of the two branches is used to define a change measure, which can be analyzed by thresholding to get the final binary change map.
Experiments are carried out on five homogeneous and heterogeneous remote sensing image datasets.
The proposed SSL approach is compared with other supervised and unsupervised state-of-the-art change detection methods.
Results demonstrate both improvements over state-of-the-art unsupervised methods and that the proposed SSL approach narrows the gap between unsupervised and supervised change detection.
\
\end{abstract}


\IEEEpeerreviewmaketitle

\section{Introduction}
\IEEEPARstart{C}{hange} maps are one of the most important products of remote sensing and are widely used in many applications including damage assessment and environmental monitoring. 
The spatial and temporal resolutions play a crucial role in obtaining accurate and timely change detection maps from multi-temporal images.
In this context, irrelevant changes, such as radiometric and atmospheric variations, seasonal changes of vegetation, and changes in the building shadows, which are typical of multi-temporal images, limit the accuracy of change maps.

In the past decades, many researchers developed techniques that directly compare pixels values of multi-temporal images to get the change maps from coarse resolution images \cite{bruzzone1997detection,bruzzone2000automatic,bovolo2006theoretical}, assuming that the spectral information of each pixel can completely characterize various underlying land-cover types.
Image rationing and change vector analysis (CVA) \cite{bruzzone2000automatic} are early examples of such algebraic approaches.
With the development of remote sensing satellite technology, the spatial and spectral resolutions of remote sensing images have significantly increased. 
In this context, the use of spectral information only is often not enough to distinguish accurately land-cover changes.
Accordingly, the joint use of spatial context and spectral information to determine the land-cover changes has gained popularity.
Many supervised \cite{bruzzone2006multilevel} and unsupervised \cite{ghosh2007context} techniques have been developed in this context.
Most of them are based on image transformation algorithms where the crucial point is to obtain robust spatial-temporal features from multi-temporal images.
Recently, deep learning techniques and in particular Convolutional Neural Networks (CNNs) methods \cite{saha2019unsupervised} have been widely used in this domain. 
CNNs allows one to get effective and robust features for the change detection tasks, achieving state-of-the-art results in a supervised way \cite{zhan2017change}.

Most of the past works are limited to the use of single modality images that are acquired by the same type of sensor with identical configurations.
Cross-domain change detection has not received sufficient attention yet. 
Current Earth Observation satellite sensors provide abundant multi-sensor and multi-modal images. 
On the one hand, images taken by different types of sensors can improve the time resolution thus satisfying the requirement of specific applications with tight constraints.
A possible example of this is the joint use of Sentinel-2 and Landsat-8 images for a regular and timely monitoring of burned areas \cite{roy2019landsat}.
However, the differences in acquisition modes and sensor parameters present a big challenge for traditional methods.
On the other hand, multimodal data are complementary to the use of single modality images and their use becomes crucial especially when only images from different sensors are available in some specific scenarios.
This could be the case of emergency management when, for example, optical and SAR images could be jointly exploited for flood change detection tasks \cite{huang2020rapid}.
In this scenario, methods capable of computing change maps from images of different sensors in the minimum possible time can be very useful. 
This has led to the development of multi-source change detection methods, which can process either multi-sensor or multi-modal images.

Recent success of deep learning techniques in change detection is mainly focused on supervised methods \cite{daudt2018fully, rahman2018siamese, peng2019end}, which are often limited from the availability of annotated datasets.
Especially in multi-temporal problems, it is expensive and often not possible to obtain a large amount of annotated samples for modeling change classes.
Thus, unsupervised methods are preferred to supervised ones in many operational applications.
The limited access to labeled data has driven the development of unsupervised methods, such as  Generative Adversarial Network (GAN)\cite{goodfellow2014generative} and Convolutional AutoEncoder (CAE)\cite{masci2011stacked}, which are currently among the most used deep learning methods in unsupervised change detection tasks.
Nevertheless, some studies have shown that such generative models overly focus on pixels rather than on abstract feature representations \cite{liu2020self}.
Recent researches in contrastive self-supervised learning \cite{tian2019contrastive,oord2018representation, grill2020bootstrap,he2020momentum} encourage the network to learn more interpretable and meaningful feature representations.
This results in improvements on classification and segmentation tasks, where they outperformed the generative counterparts.

In this work, we present an approach to perform unsupervised change detection in multi-view remote sensing images, such as multi-temporal and multi-sensor images.
The proposed approach is based on two state-of-the-art self-supervised methods, i.e., multi-view contrastive learning \cite{tian2019contrastive} and BYOL \cite{grill2020bootstrap}, that are exploited for feature representation learning.
To this purpose, a pseudo-Siamese network (which exploits ResNet-34 as the backbone) is trained to regress the output between two branches (target and online sub-networks) that were pre-trained by a contrastive way on a large archived multi-temporal or multi-sensor images dataset.
In addition, we introduce a change score that can accurately model the feature distance between bi-temporal images.
Changes are identified when there is a significant disagreement between the feature vectors of the two branches.

The rest of this paper is organized as follows.  
Section II presents the related works of unsupervised change detection in multi-view images including homogeneous and heterogeneous images.  
Section III introduces the proposed approach by describing the architecture of the pseudo-Siamese network, the two considered contrastive learning strategies and the change-detection method. 
The experimental results obtained on five different datasets and the related comparisons with supervised and unsupervised state-of-the-art methods are illustrated in Section IV. 
Finally, Section V draws the conclusions of the paper.

\section{Related Works}
In the literature, unsupervised change detection techniques in multi-view remote sensing images can be subdivided into two categories: homogeneous remote sensing image change detection and heterogeneous remote sensing image change detection.
Homogeneous image change detection methods are proposed to process multi-temporal images acquired by the same sensor or multi-sensor images with the same characteristics.
Heterogeneous image change detection methods focus on processing heterogeneous images, which are captured by different types of sensors with different imaging mechanism.

CVA \cite{bruzzone2000automatic} and its object-based variants are one of the most popular unsupervised homogeneous change detection methods.
They calculate the change intensity maps and the change direction for change detection and related classification.
Another popular method is the combination of PCA and K-means (PCA-KM)\cite{deng2008pca}, which transforms and compares the bi-temporal images in the feature space, and then determine the binary change map using k-means.
In \cite{nielsen1998multivariate}, Nilsen \textit{et al.} treated the bi-temporal images as multi-view data and proposed the multivariate alteration detection (MAD) based on canonical correlations analysis (CCA), which maximizes the correlation between the transformed features of bi-temporal images for change detection.
Wu \textit{et al.} \cite{wu2013slow} proposed a novel change detection method to project the bi-temporal images into a common feature space and detected the changed pixels by extracting the invariant components based on the theory of slow feature analysis (SFA).
As for homogeneous multi-sensor images, Solano \textit{et al.} integrated CVA into a general approach to perform change detection between multi-sensor very high resolution (VHR) remote sensing images \cite{solano2018approach}.
In \cite{ferraris2017detecting}, Ferraris \textit{et al.} introduced a CVA-based unsupervised framework for performing change detection of multi-band optical images with different spatial and spectral resolutions.

However, the traditional methods are easily affected by the irrelevant changes due to their weak feature representation ability in presence of high-resolution remote sensing images \cite{bruzzone2012novel}.
To get a robust feature representation, deep learning techniques are widely used in remote sensing change detection tasks.
In \cite{liu2019bipartite}, Liu \textit{et al.} projected the bi-temporal images into a low-dimension feature space using the restricted Boltzmann machines (RBMs) and generated change maps based on the similarity of image feature vectors.
Du \textit{et al.} \cite{du2019unsupervised} developed the slow feature analysis into deep learning methods to calculate the change intensity maps and highlight the changed components in the transformed feature space.
Then the binary change map was generated by image thresholding algorithms.
Instead of pixel-based analysis, Saha \textit{et al.} \cite{saha2019unsupervised} used a pre-trained CNNs to extract deep spatial-spectral features from multi-temporal images and analyzed the features using traditional CVA.
As an unsupervised learning method, generative models also are used in unsupervised change detection.
Lv \textit{et al.} \cite{lv2018deep} adopted a contractive autoencoder to extract features from multi-temporal images automatically. 
In \cite{ren2020unsupervised}, Ren \textit{et al.} proposed to use GAN to generate the features of unregistered image pairs and detected the changes by comparing the generated images explicitly.

Unlike homogeneous change detection, the greatest challenge in unsupervised heterogeneous change detection is to align the inconsistent feature representation of different modality images.
This requires transforming heterogeneous representation into a common feature space where performing change detection.
There are a few traditional methods that focus on this transformation of different modalities.
Gong \textit{et al.} \cite{gong2016coupled} proposed an iterative coupled dictionary learning method that learns two couple dictionaries for encoding bi-temporal images.
Luppino \textit{et al.} \cite{luppino2019unsupervised} proposed to perform image regression by transforming images to the domain of each other and to measure the affinity matrice distance, which indicates the change possibility of each pixel.
Sun \textit{et al.} \cite{sun109nonlocal} developed a nonlocal patch similarity-based method by constructing a graph for each patch and establishing a connection between heterogeneous images.

Because of the ability of CNNs in feature learning, more and more techniques based on deep learning were also proposed in this area.
Zhao \textit{et al.} \cite{zhao2017discriminative} proposed a symmetrical convolutional coupling network (SCCN) to map the discriminative features of heterogeneous images into a common feature space and generated the final change map by setting a threshold.
Similarly, the conditional generative adversarial network (cGAN) was also used to translate two heterogeneous images into a single domain  \cite{niu2018conditional}.
Luppino \textit{et al.} used the change probability from \cite{luppino2019unsupervised} as the change before to guide the training of two new networks, the X-Net with two fully convolutional networks and the adversarial cyclic encoders network (ACE-Net) with two autoencoders whose code spaces are aligned by adversarial training \cite{luppino2020deep}. 
In \cite{luppino2020code}, they further jointly used domain-specific affinity matrices and autoencoders to align the related pixels from input images and reduce the impact of changed pixels.
These methods also work well for homogeneous multi-sensor images.

\section{Methodology}
In this section, we present the proposed approach to multi-temporal and multi-sensor remote sensing image change detection based on self-supervised learning.
\subsection{Problem Statement}
Change detection is the operation of distinguishing changed and unchanged pixels of multi-temporal images acquired by different sensors at different dates.
Let us consider two images $I_1$ and $I_2$ acquired at two different dates $t_1$ and $t_2$, respectively.
The aim of change detection is to create a change intensity map that contains the most salient changed pixels, from multi-view images $I_1$ and $I_2$.
As described in related works, the crucial point in this task is to align the features of unchanged pixels or patches from the different view data $T_{1}(\theta)=f_\theta(p_1)$ and $T_{2}(\phi)=g_\phi(p_2)$.
Here, $p_1$ and $p_2$ are unchanged patches or pixels in images $I_1$ and $I_2$, respectively.
The $f$ and $g$ functions are used to extract the features from multi-temporal images, where $\theta$ and $\phi$ denote the corresponding parameters.
The objective function of our task can be defined as:
\begin{equation}\label{eq1}
\theta, \phi={\arg\min\limits_{\theta, \phi}}\{d[f_\theta(p_1),g_\phi(p_2)]\}
\end{equation}
where $d$ is a measure of feature distance between $T_1$ and $T_2$.

Many change detection techniques follow this formulation including CCA, canonical information analysis (CIA), and post-classification comparison (PCC).
CCA and CIA are used to calculate a linear/nonlinear relationship between features from multi-temporal images.
In classification-based approaches, $f$ and $g$ represent two classifiers trained independently or jointly \cite{shi2020change}.
While these change detection algorithms have made some contributions to the various application scenarios, they suffer some serious drawbacks, such as the variation in data acquisition parameters and the detection of unwanted irrelevant changes.
Thus, we still need the development of robust models, especially when the relevant changes are very hard to differentiate from the images.
With the development of deep learning, the multi-view contrastive loss and BYOL \cite{chen2016infogan} were introduced in a multi-view setting to get robust features.
These methods are considered in this work as they can extract multi-view features by maximizing the mutual information of unchanged pixels or patches between views.
In the following subsections, we will describe the proposed approach by introducing the pseudo-Siamese network, two self-supervised methods (the multi-view contrastive loss and BYOL) as well as the change detection strategy for obtaining change maps.

\subsection{Pseudo-Siamese Network}
Siamese networks \cite{bromley1993signature} are the most used model in entities comparison.
However, the comparison of heterogeneous image pairs can not be performed by Siamese networks directly for their different imaging mechanism.
Siamese networks share identical weights in two branches, while heterogeneous image pairs have dissimilar low-level features. 
Hence, the pseudo-Siamese network is used as the model architecture for heterogeneous image change detection.
It has two branches that share the same architecture except for the input channel, but with different weights.

\begin{figure*}[pt]
	\centering
	\includegraphics[width=6.5in]{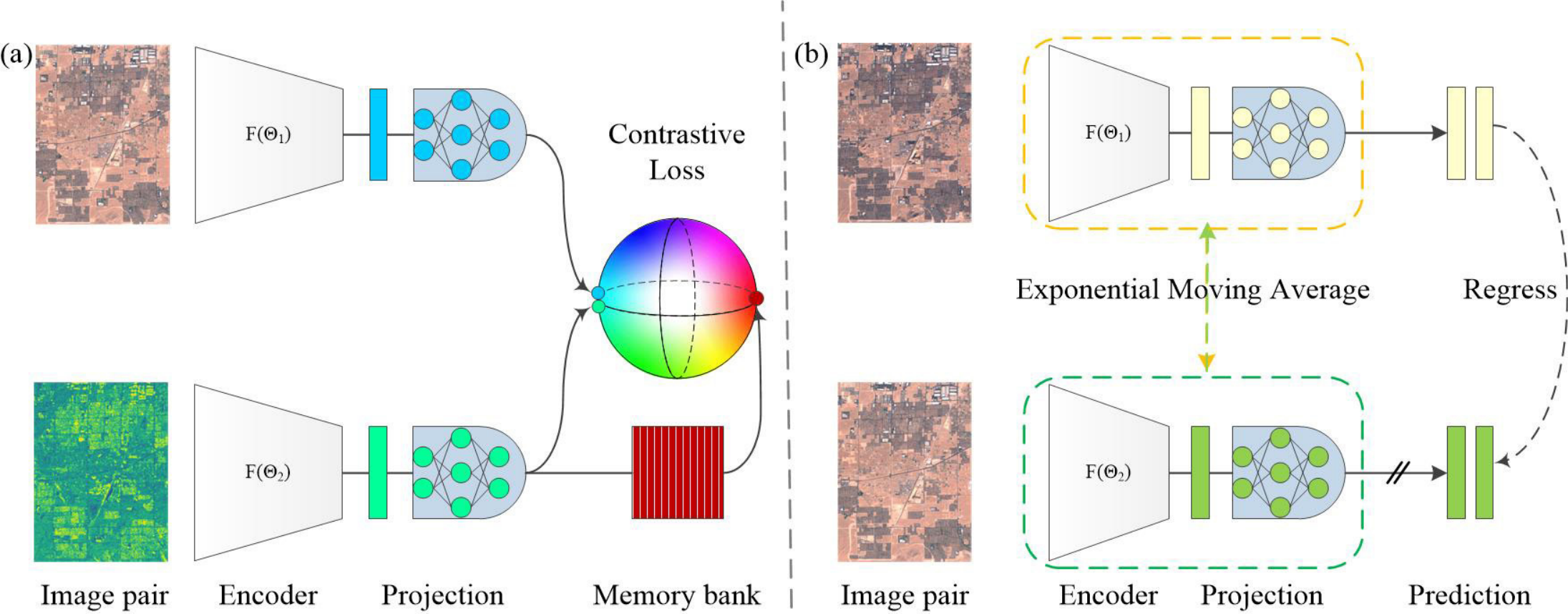}
	\caption{Pretraining part of the proposed approach to change detection (a) for heterogeneous remote sensing images and (b) for homogeneous remote sensing images. In the heterogeneous setting, the image pair consists of two images acquired by different types of sensors and the architecture of the network is symmetric with each side consisting of an encoder and a projection layer. In the homogeneous setting, the image pair consists of bi-temporal images acquired by the same sensor, and two symmetric subnetworks that share almost identical architectures but no prediction in the target subnetwork.}
	\label{fig1}
\end{figure*}

Fig. \ref{fig1} (a) shows the architecture used in this work for heterogonous change detection, where two branches are designed to extract the features of heterogeneous image pairs.
In this work, the ResNet-34 \cite{he2016deep} is adopted as the backbone of the two branches and the input channels are changed for adapting to the heterogeneous image pairs, i.e., the polarization of SAR image patches and the spectral bands of optical images patches.
In greater detail, the heterogeneous image pairs are passed through the unshared branches and are then modeled in output from the related feature vectors.
The output feature vectors of two branches are normalized and then used to compute the similarity with each other and negative samples of the batch.
Finally, the model parameters are updated by maximizing a loss function.

For homogeneous images, we propose to use the mean teacher network \cite{tarvainen2017mean} as the architecture of our model (Fig. 1 (b)).
Mean teacher is a common pseudo-Siamese network used in self-supervised learning, which uses an expositional moving average (EMA) weight to produce a more accurate model than using the same weights directly in the homogeneous images setting.
In this way, the target model has a better intermediate feature representation by aggregating the information of each step.

\subsection{Self-supervised Learning Approach}
In this subsection, we present the two considered self-supervised methods that are used in our approach to heterogeneous (Fig. 1 (a)) and homogeneous (Fig. 1 (b)) remote sensing image change detection.
\subsubsection{Multi-view Contrastive Loss (heterogeneous images)}
Contrastive learning is a popular methodology for unsupervised feature representation in the machine learning community \cite{tian2019contrastive,oord2018representation}.
The main idea behind the contrastive loss is to find a feature representation that attributes the feature distance between different samples.
For heterogeneous change detection, let us consider each heterogenous image pairs $\{I_1^i,I_2^i\}_{i=1,2,\dots,N}$ on a given scene $i$, which is considered as a positive pair sampled from the joint distribution $p(I_1^i,I_2^i)$.
Another image pair $\{I_1^i,I_2^j\}$ taken from a different scene is considered as a negative pair sampled from the product of marginals $p(I_1^i)p(I_2^j)$.
The method introduces a similarity function, $h_\theta(\cdot)$, which is used to model the feature distance between positive and negative pairs.
The pseudo-Siamese network is trained to minimize the $\mathcal{L}_{\text {contrast}}^{S}$ defined as:
\begin{equation}\label{eq3}
\mathcal{L}_{\text {contrast}}^{S}=-\underset{S}{\mathbb{E}}{\left[\log \frac{h_{\theta}(I_1^1, I_2^1)}{\sum_{j=1}^{N} h_{\theta}(I_1^1, I_2^j)}\right]}
\end{equation}
where $(I_1^1,I_2^1)$ is a positive pair sample, $(I_1^1,I_2^j|j\ge 1)$ are negative pair samples and $S=\{I_1^1, I_2^1, I_2^2, \cdots, I_2^{N-1}\}$ is a set that contains $N-1$ negative samples and one positive sample.

During the training, positive image pairs are assigned to a higher value whereas negative pairs to a lower value.
Hence, the network represents positive pairs at a close distance whereas negative pairs at a high distance.
The self-supervised method takes different augmentations of the same image as positive pairs and negative pairs sampled uniformly from the different training data.
However, such a sampling strategy for negative pairs is no longer suitable in such a case.
Robinson \textit{et al.} \cite{robinson2020contrastive} proposed an effective hard negative sampling strategy to avoid the "sampling bias" due to false-negative samples with same context information as the anchor.
With this strategy, we address the difficulty of negatives sampling in the self-supervised heterogeneous change detection task.

For heterogeneous change detection, we can construct two modalities image sets $S_1$ and $S_2$ by fixing one modality and enumerating positives and negatives from the other modality.
This allows us to define a symmetric loss as:
\begin{equation}\label{eq7}
\mathcal{L}\left(S_{1}, S_{2}\right)=\mathcal{L}_{\text {contrast}}^{S_{1}}+\mathcal{L}_{\text {contrast}}^{S_{2}}
\end{equation}
In practice, the NCE method is used to make a tractable computation of (\ref{eq7}) when $N$ is extremely large.
This multi-view contrastive learning approach makes the unsupervised heterogeneous change detection possible.

\subsubsection{Implicity Contrastive Learning (homogeneous images)}
Recently, a self-supervised framework (BYOL) was proposed that presents an implicit contrastive learning way without the requirements to have negative samples during the network training \cite{grill2020bootstrap}.
In this method, the pseudo-Siamese network, including online and target networks, is used to regress each other's output during the training.
The two networks are not fully identical.
The online network is followed by a predictor and the weights of the target network are updated by the EMA of the parameters of the online network.
Hence, the loss of the two networks can be written as the $l_2$ distance of each output:
\begin{equation}\label{eq8}
\mathcal{L} \triangleq \mathbb{E}_{\left(I_{1}, I_{2}\right)}\left[\left\|q_{w}\left(f_{\theta}\left(I_{1}\right)\right)-f_{\phi}\left(I_{2}\right)\right\|_{2}^{2}\right]
\end{equation}
Similar to the multi-view contrastive loss, the feature vectors are $l_2$-normalized before output.
Here the online network $f_\theta$ is parameterized by $\theta$, and $q_w$ is the predictor network parameterized by $w$. 
The target network $f_{\phi}$ has the same architecture as $f_{\theta}$ but without the final predictor and its parameters are updated by EMA controlled by $\tau$, i.e.,
\begin{equation}\label{eq9}
\phi \leftarrow \tau \phi+(1-\tau) \theta
\end{equation}

The most important property of BYOL is that no negative samples are used when training the two networks, and thus feature representations are learned only from positive samples.
A previous work \cite{chen2020exploring} has pointed out that the architecture of Siamese network is the key to implicit contrastive learning and the predictor with batch-normalization can avoid the representation collapse during the training.
In this approach, the network is identical in the two branches, and the weights of the target part are updated according to another branch.
Hence, this algorithm is very suitable to process multi-temporal remote sensing images with the same modality (i.e., homogeneous images).

\begin{figure*}[pt]
	\centering
	\includegraphics[width=6.5in]{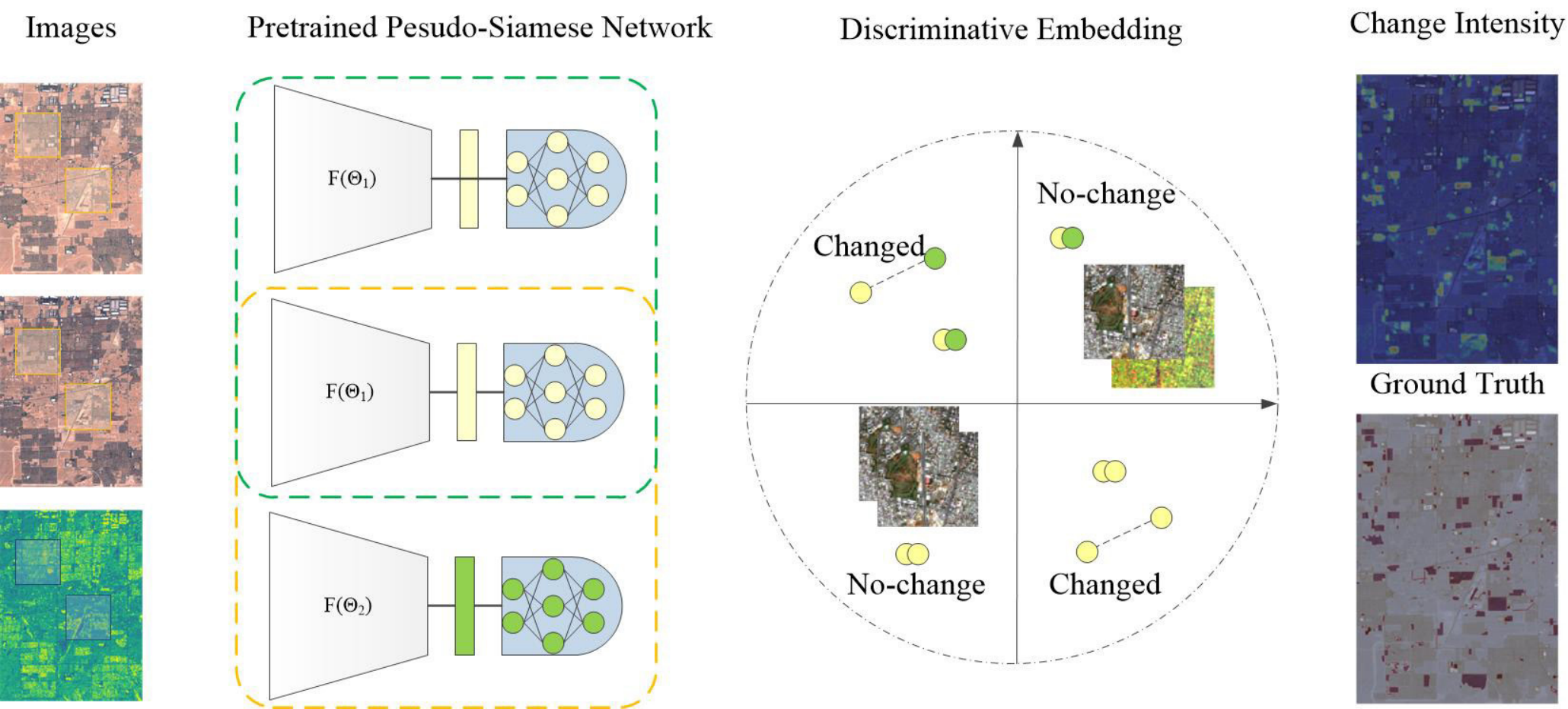}
	\caption{Schematic overview of the proposed change detection approach (SSL). Input images are fed through the pre-trained pseudo-Siamese network that extracts feature vectors from homogeneous or heterogeneous bi-temporal image patches. Then, the pre-trained pseudo-Siamese network estimates regression errors for each pixel. Change intensity maps are generated by combining results with a different patch side length and the final binary change map is obtained by setting a threshold.}
	\label{fig2}
\end{figure*}
\subsection{Change Detection}
The change detection strategy described in this subsection is based on the feature learned by the previously mentioned self-supervised methods.
Let $S=\{I_1, I_2, I_3, ..., I_n\}$ be a dataset of either homogeneous or heterogeneous multi-temporal remote sensing images.
Our goal is to detect changes between satellite images from different dates.
As mentioned before, most changes of interest are those relevant to human activities, while the results are easily affected by irrelevant changes, such as seasonal changes.
Other relevant changes are usually rare, whereas irrelevant changes are common during a long period.
This means that, under this assumption, the features of relevant changes can be derived from the unchanged features.
To this purpose, the models are trained to regress the features of images acquired at different dates.
As shown in Fig. \ref{fig2}, here we use the considered self-supervised learning algorithms to get features of either homogeneous or heterogeneous multi-temporal images.
After training, a change intensity map can be derived by assigning a score to each pixel indicating the probability of change.

During the network training, images acquired by the different sensors or at different dates are treated as two-views in our approach.
Homogeneous images are trained with BYOL, while heterogeneous images are trained by using multi-view contrastive loss.
Image patches centered at each pixel are fed in input to the network, and the output is a single feature vector for each patch-sized input.
In detail, given an input image $\mathbf{I} \in \mathbb{R}^{w \times h}$ of width $w$, height $h$, we can get a feature vector $T(r,c)$ of a square local image region  with a side length $p$ for each image pixel at row $r$ and column $c$.
To get different scale feature representations, we trained an ensemble of $N \geq 1$ randomly initialized models that have an identical network architecture but use different input image sizes.
Therefore, changes of different sizes are detected by choosing one of the $N$ different side length values.
During the inference, each model provides as output a feature map that is generated by different sizes of input images.
Let ${T}_1^{i}(r,c)$ and ${T}_2^{i}(r,c)$ denote the feature vectors at the row $r$ and column $c$ for the considered bi-temporal images.
The change intensity map is defined as the pair-wise regression error $e(r,c)$ between the feature vectors of bi-temporal images:
\begin{equation}
\begin{aligned}
e{(r, c)} &=\left\|T_1{(r, c)}-T_2{(r, c)}\right\|_{2}^{2} \\
&=\left\|\frac{1}{N} \sum_{i=1}^{N} \left({T}_1^{i}(r,c)-{T}_2^{i}(r,c)\right)\right\|_{2}^{2}
\end{aligned}
\end{equation}
In order to allow all model outputs to be merged, we normalize each output by its mean value $e_\mu$ and standard deviation $e_\sigma$.
Therefore, multi-scale change detection can be simplified into sub-tasks that train multiple pseudo-Siamese ensemble networks with varying values of $p$.
At each scale, a change intensity map with the same size as the input image is computed.
Given $N$ pseudo-Siamese ensemble models with different side length, the normalized regression error $\tilde{e}(r, c)$ of each model can be combined by simple averaging.

One can see from Fig. \ref{fig2} that pixels can be classified as changed and unchanged by thresholding the feature distance in the change intensity map.
In this case, two strategies are considered.
The simplest strategy is to choose the opposite minimum value of standardized intensity maps as the threshold value.
An alternative strategy is the Robin thresholding method \cite{rosin2005remote}, which is robust and suitable for long-tailed distribution curves.
In this method, the threshold value is the "corner" on the distribution curve of the intensity map and the maximum deviation from the straight line drawn between the endpoints of the curve.
In our technique, the threshold value is determined by the first strategy if the absolute difference of these two threshold values is smaller than half of their average value.
Otherwise, the threshold value is determined by the Robin thresholding method.

\section{Experimental Results}
In this section, we first present the considered datasets, then the state-of-the-art change detection methods used in the comparison, and finally conduct a thorough analysis of the performance of different approaches and of their results.
\subsection{Description of Datasets}
We developed our experiments on five different datasets including three homogeneous datasets and two heterogeneous datasets. 
All remote sensing images in this work are raw images from the google earth engine (GEE) and without any specific pre-processing.
\subsubsection{OSCD\_S2S2/\_S1S1/\_S1S2/\_L8S2}
The Onera Satellite Change Detection (OSCD) dataset \cite{daudt2018urban} was created for bi-temporal change detection using Sentinel-2 images acquired between 2015 and 2018.
These images have a total of 13 bands with a relatively high resolution (10 m) for Visible (VIS) and near-infrared (NIR) band images and 60 m resolution for other spectral channels.
The images of this dataset include urban areas and present the change type of urban growth and changes.
The dataset consists of 24 pairs of multispectral images and the corresponding pixel-wise ground truth acquired in different cities and including different landscapes.
The pixel-wise ground truth labels, which were manually annotated, were also provided for each pair but with some errors due to the relatively limited resolution of Sentinel-2 images.
At the original supervised setting, 14 pairs were selected for the training set and the rest 10 pairs were used to evaluate the performance of methods.

To use this dataset in self-supervised training, we downloaded additional Sentinel-2 images in the same location as the original bi-temporal images between 2016 and 2020.
We considered images from each month to augment existing image pairs.
Similarly, Landsat-8 multi-temporal images and Sentinel-1 ground range detected (GRD) image products are also provided in this dataset corresponding to the given Sentinel-2 scenes.
The Landsat-8 images have nine channels covering the spectrum from deep blue to shortwave infrared and two long-wave infrared channels and their resolution range from 15 m to 100 m.
The Sentinel-1 GRD products have been terrain corrected, multi-looked, and transformed to the ground range and geographical coordinates. 
They consist of two channels including Vertical-Horizontal (VH) and Vertical-Vertical (VV) polarization as well as of additional information on the incidence angle.

To use this dataset for multi-view change detection, we separate it into four sub-datasets: OSCD{\_}S2S2, OSCD{\_}S1S1, OSCD{\_}S1S2 and OSCD{\_}L8S2.
These datasets are composed of homogeneous multi-temporal optical or SAR images (OSCD\_S2S2, OSCD\_S1S1, OSCD\_L8S2) and heterogeneous multi-temporal SAR-optical images (OSCD\_S1S2).
To keep consistency with previous research, 10 image pairs of these four datasets corresponding to the OSCD test image pairs are treated as the test dataset to evaluate the performance of different methods, and image pairs acquired on other scenes and on each month of four years are used for the self-supervised pre-training.
In practice, it is impossible to acquire the test image pairs of OSCD\_S1S1, OSCD\_L8S2, and OSCD\_S1S2 at the same time as the OSCD\_S2S2.
Hence, we only obtained these image pairs at the closest time to OSCD\_S2S2 test image pairs.

\subsubsection{Flood in California}
The California dataset is also a heterogeneous dataset that includes a Landsat-8 (multi-spectral) and a Sentinel-1 GRD (SAR) image.
The multispectral and SAR images are acquired on 5 January 2017 and 18 February 2017, respectively.
The dataset represents a flood occurred in Sacramento County, Yuba County, and Sutter County, California.
The ground truth was extracted from a Sentinel-1 SAR image pair where the pre-event image is acquired approximately at the same time as the Landsat-8 image.
However, we realized that the ground truth in \cite{luppino2019unsupervised} contains many mistakes.
Hence, we updated the reference data with the PCC method according to bi-temporal Sentinel-1 images.
Other three image pairs of Sentinel-1 and Landsat-8 images of the same scene acquired in 2017 and 2018, respectively, were used for the self-supervised pre-training of the proposed SSL approach.

\subsection{Experimental Settings}

\subsubsection{Literature Methods for Comparison}
We considered different state-of-the-art methods for comparisons with the proposed SSL approach on the five datasets mentioned above.
On the first two homogeneous datasets (OSCD\_S2S2 and OSCD\_L8S2), the proposed SSL approach was compared with two unsupervised deep learning approaches (DSFA \cite{du2019unsupervised} and CAA \cite{luppino2020code}) and two deep supervised methods (FC-EF \cite{daudt2018fully} and FC-EF-Res \cite{daudt2019multitask}).

Deep Slow Feature Analysis (DSFA) is a deep learning-based multi-temporal change detection method consisting of two symmetric deep networks and based on the slow feature analysis theory (SFA).
The two-stream CNNs are used to extract image features and detect changes based on SFA.
Code-Aligned Autoencoders (CAA) is a deep unsupervised methodology to align the code spaces of two autoencoders based on affinity information extracted from the multi-modal input data. 
It allows achieving a latent space entanglement even when the input images contain changes by decreasing the interference of changed pixels.
However, it degrades its performance when only one input channel is considered.
It is also well suited for homogeneous change detection, as it does not depend on any prior knowledge of the data.

Fully convolutional-early fusion (FC-EF) is considered for the supervised change detection method on the OSCD dataset. 
In this method, the bi-temporal image pair are stacked together as the input. 
The architecture of FC-EF is based on U-Net \cite{ronneberger2015u}, where the skip connections between encoder and decoder help to localize the spatial information more precisely and get clear change boundaries.
FC-EF-Res is an extension of FC-EF with residual blocks to improve the accuracy of change results.
In addition, it is worth noting that the first dataset (OSCD\_S2S2) has previously been extensively used in other works.
Hence, we also compare our results with those of some conventional methods \cite{daudt2018urban} (Log-ratio, GLRT and Image difference), an unsupervised deep learning method (ACGAN \cite{saha2019unsupervised2}) and supervised deep learning techniques (FC-Siam-conc and FC-Siam-diff \cite{daudt2018urban}) reported in previous papers.

On the Sentinel-1 SAR images dataset, only unsupervised methods (DSFA, SCCN, and CAA) are used for comparison.
Note that some change information present in multi-spectral images is not detectable in SAR images, hence we did not use supervised methods on them.
On the two heterogeneous remote sensing image datasets (OSCD\_S1S2 and California),  two state-of-the-art methods are used for comparisons, including the symmetric convolutional coupling network (SCCN) and CAA.
Considering that only significant changes in the backscattering of SAR images can be detected, we only consider the LasVegas site in the OSCD\_S1S2 dataset.
Similar to CAA, SCCN is an unsupervised multi-modal change detection method that exploits an asymmetrical convolutional coupling network to project the heterogeneous image pairs onto the common feature space.
This method is also used in the homogeneous SAR image pairs in our experiments.

\subsubsection{Implementation details}
We take the ResNet-34 as the backbone of two branches of the pseudo-Siamese network to get feature vectors of corresponding image patches.
In particular, we change the parameters of the strider from 2 to 1 in the third and fourth layers of the backbone for adapting the network to the relatively small input size.
In order to capture the different scales of change, we use three different patch sizes (p = 8, 16, 24 pixels) for the homogeneous image change detection task and two different patch sizes (p = 8, 16 pixels) for the heterogeneous change detection task. 

During the training on OSCD\_S2S2, we randomly composed all images acquired at different dates into pairs as the input.
While SAR/multi-spectral image pairs acquired in the same month have been used as input pairs for the rest of the multi-sensor dataset.
After finishing the training process, the test image pairs are feed into the pre-trained network and then the related change intensity maps are derived.   
For the supervised method (FC-EF and FC-EF-Res), we used the 14 bi-temporal training images considered in the previous work \cite{daudt2019multitask}. 
In the self-supervised and supervised method, we use four channels (VIS and NIR) in Landsat-8 and Sentinel-2 images, while two polarizations (VH and VV) in Sentinel-1 images.
CAA and SCCN methods require heterogeneous image pairs having the same number of input channels.
According, to keep consistency with the four input channels of multi-spectral images, we augmented Sentinel-1 images with the plus and minus operation between the two polarizations as the other two channels.

\subsubsection{Evaluation Criteria}
To appraise the different methods presented above, five evaluation metrics (precision (Pre), recall (Rec), overall accuracy (OA), F1 score and Cohen's kappa score (Kap)) are used in this paper.
We simply classify the image pixels into two classes by setting an appropriate threshold value according to the presented strategy and analyze them with reference to the ground truth map.
Then, the number of unchanged pixels incorrectly flagged as change is denoted by $FP$ (false positive) and the number of changed pixels incorrectly flagged as unchanged is denoted by $FN$ (false negative).
In addition, the number of changed pixels correctly detected as change is denoted by $TP$ (true positive) and the number of unchanged pixels correctly detected as unchanged is denoted by $TN$ (true negative).
From these four quantities, the five evaluation metrics can be defined as :
\begin{equation}\label{eq21}
Pre=\frac{TP}{TP+FP} 
\end{equation}
\begin{equation}\label{eq22}
Rec=\frac{TP}{TP+FN}
\end{equation}
\begin{equation}\label{eq23}
F_{1}=\frac{2 Pre \cdot Rec}{Pre+Rec}
\end{equation}
\begin{equation}\label{eq24}
OA=\frac{TP+TN}{TP+TN+FP+FN}
\end{equation}
\begin{equation}\label{eq25}
Kap=\frac{OA-PE}{1-PE}
\end{equation}
\begin{equation}\label{eq26}
\begin{aligned}
\mathrm{PE}&=\frac{(TP+FP) \cdot(TP+FN)}{(TP+TN+FP+FN)^{2}}\\
&+\frac{(FN+TN) \cdot(FP+TN)}{(TP+TN+FP+FN)^{2}}
\end{aligned}
\end{equation}
Obviously, a higher value of $Pre$ results in fewer false alarms, and a higher value of $Rec$ represents a smaller rate of incorrect detections.
The overall accuracy $OA$ is the ratio between correctly detected pixels and all pixels of the image.
However, these three metrics will give a misleading over-estimate of the result when the amount of changed pixels is a small fraction of the image.
$F1$ score and $Kap$ can overcome the problem of $Pre$ and $Rec$ and better reveal the overall performance.
Note that large $F1$ and $Kap$ values represent better overall performance.

\begin{figure*}[pt]
	\centering
	\includegraphics[width=6.5in]{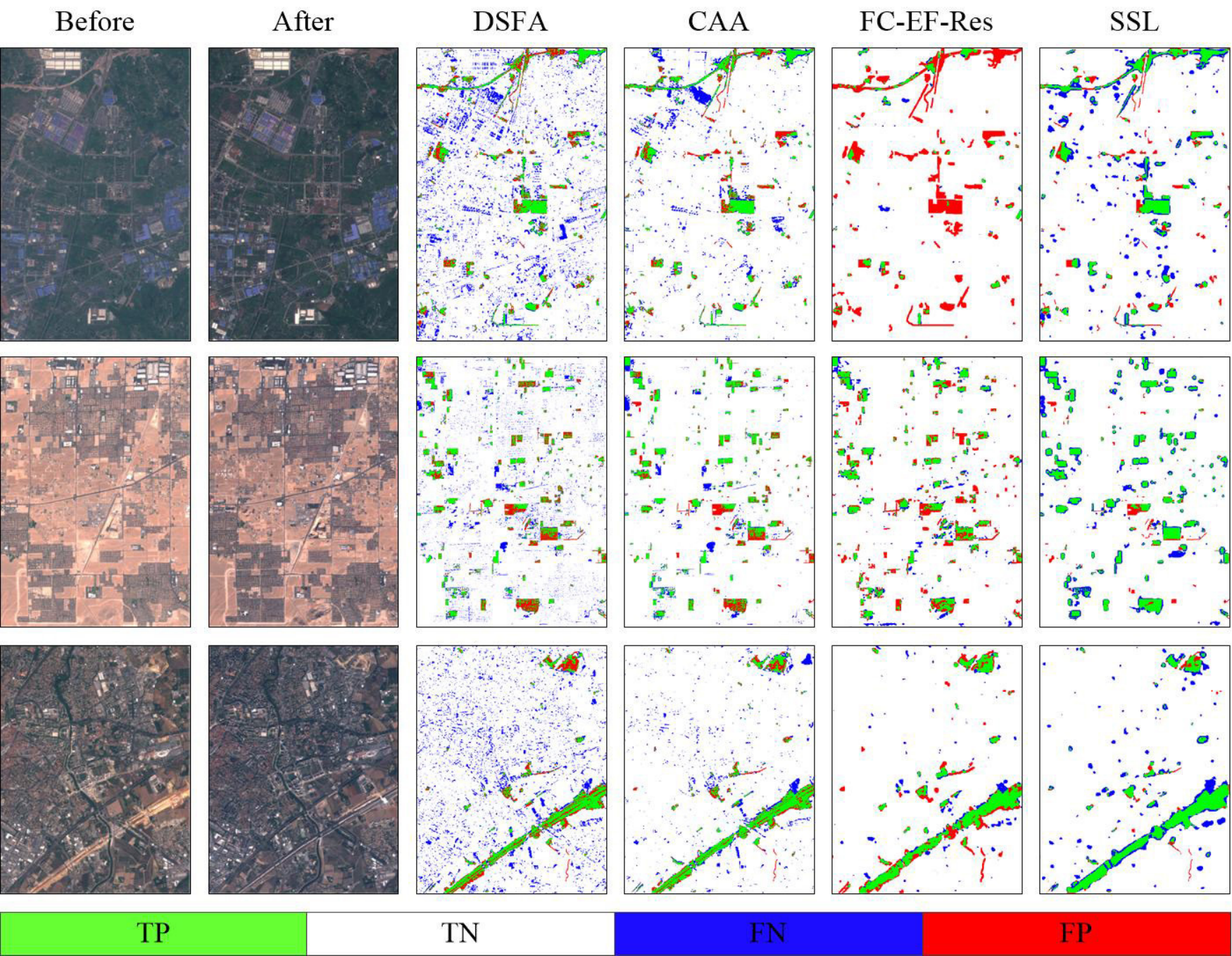}
	\caption{Examples of change detection results on OSCD\_S2S2, organized in one row for each location. Col. 1: pre-event image (Sentinel-2); Col. 2: post-event image (Sentinel-2). Change maps obtained by: DSFA (Col. 3), CAA (Col. 4), FC-EF-Res (Col. 5), and the proposed SSL (Col. 6).}
	\label{fig3}
\end{figure*}

\begin{figure*}[pt]
	\centering
	\includegraphics[width=6.5in]{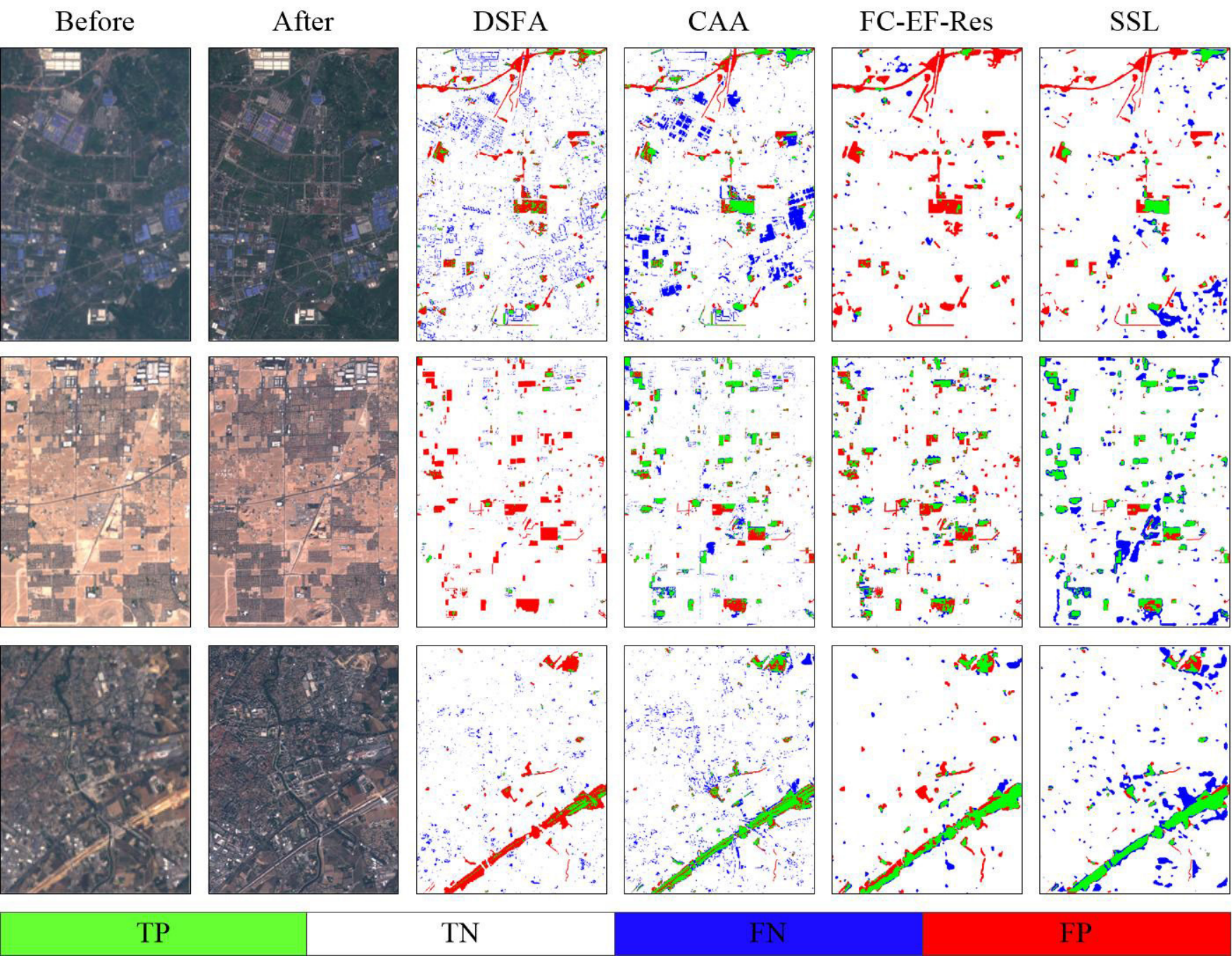}
	\caption{Examples of change detection results on OSCD\_L8S2, organized in one row for each location. Col. 1: pre-event image (Landsat-8); Col. 2: post-event image (Sentinel-2). Change maps obtained by: DSFA (Col. 3), CAA (Col. 4), FC-EF-Res (Col. 5), and the proposed SSL (Col. 6). }
	\label{fig4}
\end{figure*}

\begin{figure*}[pt]
	\centering
	\includegraphics[width=6.5in]{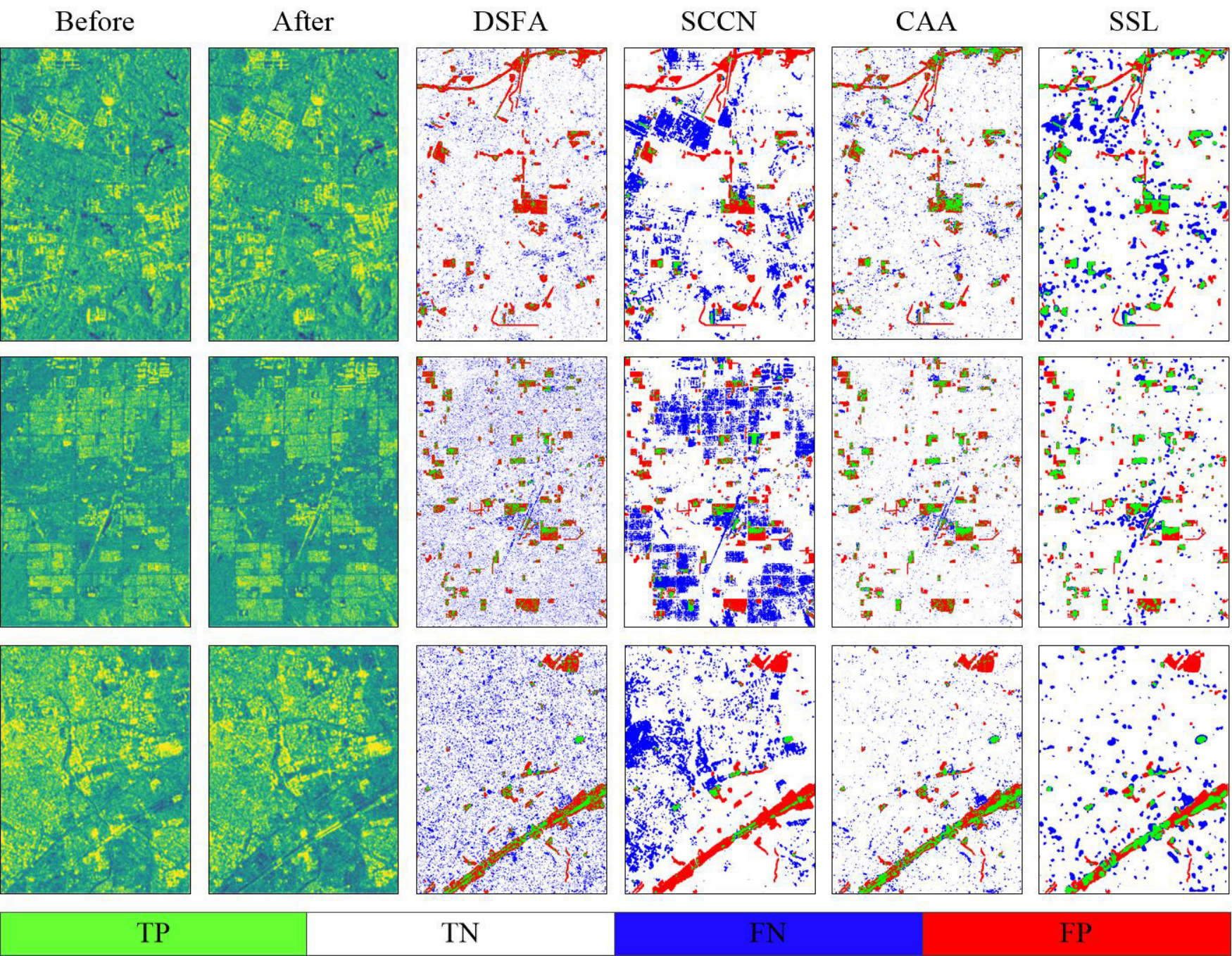}
	\caption{Examples of change detection results on OSCD\_S1S1, organized in one row for each location. Col. 1: pre-event image (Sentine-1); Col. 2: post-event image (Sentine-1). Change maps obtained by: DSFA (Col. 3), SCCN (Col. 4), CAA (Col. 5), and the proposed SSL (Col. 6). }
	\label{fig5}
\end{figure*}

\subsection{Results on Homogeneous Datasets}
\begin{table}[pb]
	\centering
	\caption{Quantitative evaluations of different approaches applied to the homogeneous images OSCD\_S2S2 dataset.}
	\label{table1}
	\renewcommand\tabcolsep{5.0pt} 
	\centering
	\begin{tabular}{ccccccc}
		\hline
		Type & Method & Pre(\%) & Rec(\%) & OA(\%) & F1 & Kap \\ \hline
		\multirow{7}{*}{\rotatebox{90}{Unsupervised}} & Prop. SSL & 36.95 & 59.48 & 92.50 & 0.46 & 0.42 \\
		& DSFA & 26.77 & 54.24 & 92.63 & 0.36 & 0.32 \\
		& AAC & 23.49 & 52.96 & 91.66 & 0.33 & 0.29 \\	
		& ACGAN[44] & - & 64.63 & 77.67 & - & - \\
		& Img. Diff[41] & - & 63.42 & 76.12 & - & - \\
		& GLRT[41] & - & 60.48 & 76.25 & - & - \\
		& Log-ratio[41] & - & 59.68 & 76.93 & - & - \\ \hline
		\multirow{8}{*}{\rotatebox{90}{Supervised}} & FC-EF & 55.34 & 39.48 & 95.13 & 0.46 & 0.44 \\
		& FC-EF-res & 54.97 & 38.39 & 95.10 & 0.45 & 0.43 \\
		& Siamese[41] & 21.57 & 79.40 & 76.76 & 0.34 & - \\
		& EF[41] & 21.56 & 82.14 & 83.63 & 0.34 & - \\
		& FC-EF*[42] & 44.72 & 53.92 & 94.23 & 0.49 & - \\
		& FC-EF-Res*[42] & 52.27 & 68.24 & 95.34 & 0.59 & - \\
		& FC-Siamese-Con*[42] & 42.89 & 47.77 & 94.07 & 0.45 & - \\
		& FC-Siamese-Diff*[42] & 49.81 & 47.94 & 94.86 & 0.49 & - \\ \hline
	\end{tabular}
\end{table}
We first evaluate the change detection performance of the proposed approach and state-of-the-art methods (DSFA, CAA and supervised methods) applied to the homogeneous change detection scenario. 
This includes bi-temporal Sentinel-2 images (OSCD\_S2S2 test dataset), bi-temporal landsat-8/Sentinel-2 images (OSCD\_L8S2 test dataset) and bi-temporal Sentinel-1 images (OSCD\_S1S1 test dataset).
The performance metrics obtained on the OSCD\_S2S2 test dataset are reported in Table \ref{table1}.
As expected the FC-EF and FC-EF-Res supervised methods applied to raw images achieved the best performance in  terms of Precision, OA, F1 and Kappa, but not on Recall.
Among all unsupervised methods, the proposed SSL approach with an OA of 92.5 \% and a Kappa coefficient of 0.42, obtained the best performance on all five metrics and the third-best performance among all methods (included the supervised ones) implemented in this work.
Although two supervised methods performed better than other methods on most metrics, they have a much worse performance on Recall than the proposed SSL approach.
It is also worth noting that the proposed SSL approach is effective in closing the gap with the supervised methods on Kappa, which indicates its effective overall performance.
In addition, the results of other unsupervised methods (i.e., ACGAN, Image difference, GLRT, and Log-ratio) and supervised methods (i.e., Siamese and EF) on VIS and NIR channels in \cite{daudt2018urban} are reported in the table.
They are all worse than those of the proposed SSL approach.
The results of other supervised methods (i.e., FC-EF*, FC-EF-Res*, FC-Siamese-Con* and FC-Siamese-Diff*) applied to carefully processed RGB channel images are reported in the last rows of Table \ref{table1}.
Their accuracies on most metrics are slightly better than those of the proposed SSL approach, but they can not be achieved when working on raw images as a high registration precision is required.
Indeed, in the related papers, multi-temporal images are carefully co-registered using GEFolki toolbox to improve the accuracy of change maps \cite{daudt2018urban}.
On the contrary, the proposed SSL approach is based image patches where the registration precision of Sentinel system is enough for obtaining a good change map.

Besides the quantitative analysis, we also provide a visual qualitative comparison in Fig. \ref{fig3}, where the TP, TN, FN and FP pixels are colored in green, white, blue and red, respectively.
One can see that change maps provided by DSFA and CAA are affected by a significant salt-and-pepper noise where plenty of unchanged buildings are misclassified as changed ones.
This is due to the lack of use of spatial context information in these methods.
This issue is well addressed by the proposed SSL approach and the FC-EF-Res supervised method, which provide better maps.
Most of the changed pixels are correctly detected in the proposed SSL approach, but with more false alarms than in the supervised FC-EF-Res method.
Note that this is probably due to some small changes that are ignored in the ground truth.
Nonetheless, since these results are processed in patches, some small objects are not classified correctly and false alarms on boundaries of buildings are provided by the proposed SSL approach.
A possible reason for this is the small patch-based method with a poor spatial context information learning ability.
Instead, the change maps obtained by the FC-EF-Res method are in general more accurate and less noisy due to the use of spatial-spectral information in U-Net and the supervised learning algorithm.
However, the FC-EF-Res method failed to detect most of changed pixels in the first scenario.
This confirms that the change detection results of supervised methods heavily rely on the change type distribution and the quality of training samples.
This is not an issue for the proposed SSL approach.

The performance of each model is also validated on the OSCD\_L8S2 test dataset, which was obtained by different optical sensors having different spatial resolutions, and the quantitive evaluation is reported in Table \ref{table2}.
In general, the supervised methods outperform DSFA and CAA considering all five metrics.
However, the performance of FC-EF-res on Recall is much worse than those of CAA and the proposed SSL approach.
Meanwhile, the proposed SSL approach with an overall accuracy of 92.6\% and a Kappa coefficient of 0.29, obtained the best accuracy among other unsupervised methods and is very close to the supervised methods on all five metrics.
Fig. \ref{fig4} presents the binary change maps obtained by all methods on the OSCD\_L8S2.
One can see that the change maps contain a larger number of false alarms for all methods compared with the maps obtained on the OSCD\_S2S2.
This is probably due to the relatively lower resolution of Landsat-8 VIS and NIR channel images with respect to the counterparts in Sentinel-2 images.
Consistently with the results obtained on OSCD\_S2S2 (see Fig. \ref{fig3}), the proposed SSL approach has a better segmentation result but with lower accuracy on all metrics, which indicates that the different resolution images increase the difficulty of change detection tasks.

\begin{table}[pt]
	\centering
	\caption{Quantitative evaluations of different approaches applied to the homogeneous images OSCD\_L8S2 and OSCD\_S1S1 datasets.}
	\label{table2}
	\renewcommand\tabcolsep{5.0pt} 
	\centering
	\begin{tabular}{cccccccc}
		\hline
		Dataset & Type & Method & Pre(\%) & Rec(\%) & OA(\%) & F1 & Kap \\ \hline
		\multirow{5}{*}{L8S2} & \multirow{3}{*}{\rotatebox{90}{Unsup.}} & Prop. SSL & 31.67 & 34.59 & 92.61 & 0.33 & 0.29 \\
		& & CAA & 18.45 & 45.80 & 90.25 & 0.26 & 0.22 \\
		& & DSFA & 8.08 & 24.29 & 86.64 & 0.12 & 0.07 \\ \cline{2-8} 
		& \multirow{2}{*}{\rotatebox{90}{Sup.}} & FC-EF & 29.75 & 34.08 & 92.27 & 0.32 & 0.28 \\
		& & FC-EF-res & 39.14 & 27.14 & 93.93 & 0.32 & 0.29 \\ \hline
		\multirow{4}{*}{S1S1} & \multirow{4}{*}{\rotatebox{90}{Unsup.}} & Prop. SSL & 23.06 & 40.39 & 89.74 & 0.29 & 0.24 \\
		& & SCCN & 7.48 & 27.80 & 78.04 & 0.12 & 0.04 \\
		& & CAA & 19.80 & 34.81 & 89.12 & 0.25 & 0.20 \\
		& & DSFA & 10.96 & 22.78 & 92.63 & 0.15 & 0.08 \\ \hline
	\end{tabular}
\end{table}
To complete the evaluation on homogeneous datasets, the performance of all unsupervised methods are validated on the OSCD\_S1S1 test dataset.
The quantitative results are reported in Table \ref{table2}, which shows that the proposed SSL approach produces a better accuracy than other methods on all metrics, except for OA.
The binary change maps obtained by each unsupervised methods are shown in Fig. \ref{fig5}.
One can see that all results appear much noisier due to the influence of speckle in SAR images.
It is worth noting that only a new building that appeared in the post-event SAR image can be detected because minor growth of the building does not cause significant backscatter change.
Apart from this, the boundaries of the detected objects are not accurate as those in the optical dataset due to the side-looking imaging mechanism.
In addition, the performance of the proposed SSL approach on OSCD\_S1S1 is close to that obtained on OSCD\_L8S2 but with fewer correct detections and more false alarms than the latter.
In general, the above three experiments based on homogeneous images demonstrate that the proposed SSL approach obtained the best quantitative and qualitative performance with respect to all the other considered unsupervised change detection techniques.

\begin{figure*}[pt]
	\centering
	\includegraphics[width=6.5in]{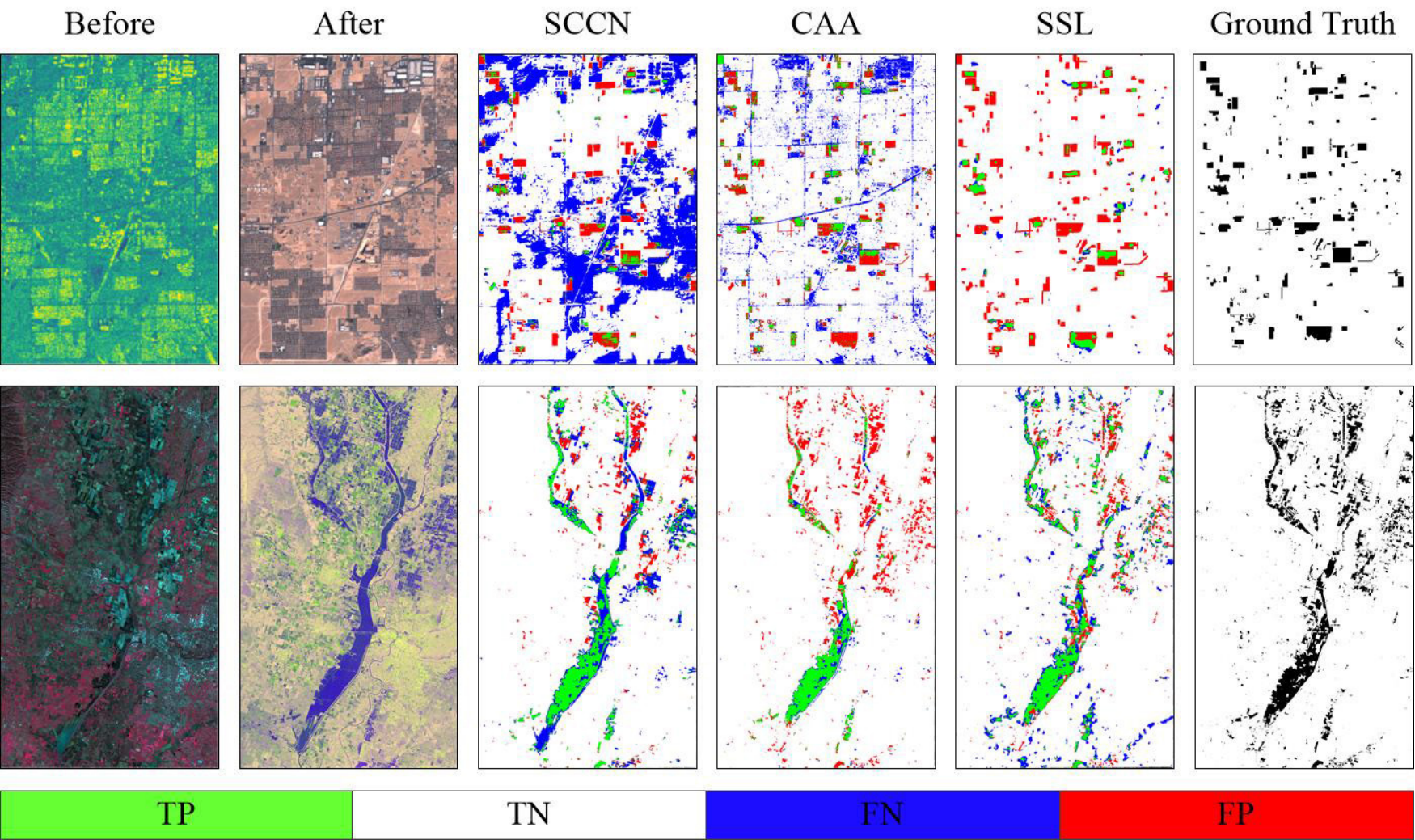}
	\caption{Change detection results on OSCD\_S1S2 and califorlia flood, organized in one row for each location. Col. 1: pre-event image (Sentine-1 for OSCD\_S1S2 and Landsat-8 for the califorlia flood); Col. 2: post-event image (Sentine-2 for OSCD\_S1S2 and Sentine-1 for the califorlia flood). Change maps obtained by: SCCN (Col. 3), CAA (Col. 4), and the proposed SSL (Col. 5). Col. 6: the ground truth. }
	\label{fig6}
\end{figure*}
\subsection{Results on Heterogeneous Datasets}
In the second change detetcion scenario, we consider two heterogeneous datasets which consist of a Sentinel-1/Sentinel-2 image pair (OSCD\_S1S2) and a Sentinel-1/Landsat-8 image pair (California).

The performance of three unsupervised methods (SCCN, CAA and SSL) on OSCD\_S1S2 is reported in Table \ref{table3}.
One can see that the proposed SSL approach performs much better than the other two unsupervised methods on most metrics due to the separated training on the archived images.
In contrast, SCCN and CAA are both trained on the test image only and the complicated background in the scene makes them hard to separate the unchanged pixels for the network training causing too many false alarms in change detection maps.
Compared with the results obtained in the homogeneous experiments, the results presented here are much worse.
This demonstrates the difficulty of heterogeneous change detection in complicated backgrounds, such as an urban area.
Fig. \ref{fig6} presents the qualitative visual results in terms of binary change maps.
One can observe that the results provided by SCCN and CAA are affected by many more missed detections and false alarms than in the homogeneous case.
The result of the proposed SSL approach has fewer false alarms but with more missed detections with respect to the homogeneous setting owing to the larger domain discrepancy.

\begin{table}[pt]
	\centering
	\caption{Quantitative evaluations of different approaches applied to the heterogeneous images OSCD\_S1S2 and the California datasets.}
	\label{table3}
	\centering
	\begin{tabular}{ccccccc}
		\hline
		Dataset & Method & Prec(\%) & Rec(\%) & OA(\%) & F1 & Kap \\ \hline
		\multirow{3}{*}{S1S2} & SCCN & 7.38 & 22.45 & 68.54 & 0.11 & - \\
		& CAA & 21.91 & 28.71 & 84.79 & 0.25 & 0.17 \\
		& Prop. SSL & 70.32 & 19.01 & 92.20 & 0.30 & 0.27 \\ \hline
		\multirow{3}{*}{California} & SCCN & 51.42 & 64.44 & 92.88 & 0.57 & 0.53 \\
		& CAA & 76.49 & 40.38 & 94.68 & 0.53 & 0.50 \\
		& Prop. SSL & 48.79 & 63.82 & 92.39 & 0.55 & 0.51 \\ \hline
	\end{tabular}
\end{table}

Differently from the previous dataset, the California dataset is related to a simpler background and to more significant changes resulted from the flood.
Table \ref{table3} presents the results of all methods on this dataset.
The three unsupervised methods (SCCN, CAA and SSL) have similar performance on overall evaluation metrics (OA, F1 and Kappa).
The SCCN achieves the best Recall, F1 score, Kappa and the second-best values on Precision and OA, while the CAA achieved the highest Precision and OA values.
The proposed SSL approach gets the second-best values on three of five metrics, thus it does not show obvious superiority.
Fig. \ref{fig6} illustrates the Landsat 8 and Sentinel-1 images and the change maps from the compared methods.
Maps provided by SCCN and ACC show a clear boundary of change areas, whereas the one of the proposed SSL approach is less precise.
The map of SCCN contains more false alarms, while the map of the CAA has more missed detections.
Even if the performance of the proposed SSL approach on the California dataset is not the best, it is still no worse than that of the other two methods considering all five metrics. 
In general, considering the results on the two heterogeneous test datasets, the proposed SSL approach is the most accurate followed by the CAA, which is the second-best method and is only slightly worse than the proposed SSL approach.

\section{Conclusion}
In this work, we have presented a self-supervised approach to unsupervised change detection in multi-view remote sensing images, which can be used with both multi-sensor and multi-temporal images.
The main idea of the presented framework is to extract a good feature representation space from homogeneous and heterogeneous images using contrastive learning.
Images from satellite mission archives are used to train the pseudo-Siamese network without using any label.
Under the reasonable assumption that the change event is rare in long-time archived images, the network can properly align the features learned from images obtained at different times even when they contain changes.
After completing the pre-training process, the regression error of image patches captured from bi-temporal images can be used as a change score to indicate the change probability.
If required, a binary change map can be directly calculated from change intensity maps by using a thresholding method.

Experimental results on both homogeneous and heterogeneous remote sensing image datasets proved that the proposed SSL approach can be applicable in practice, and demonstrated its superiority over several state-of-the-art unsupervised methods.
Results also show that the performance declines when the resolution of the two sensors is different in a homogeneous setting.
Moreover, in the SAR-optical change detection setting, the change detection results are affected by the complexity of the background.

As a final remark, note that in this work we only considered bi-temporal images to detect changes.
This has negative impacts on false alarms.
Our future work will be focused on the refinement of changed maps by further decreasing false alarms by combining a larger number of images from the time-series.

\section*{Acknowledgment}
The authors would like to thank Yuanlong Tian and Thalles Silva for their open-source code in their work. This study was supported by the China Scholarship Council.

\ifCLASSOPTIONcaptionsoff
  \newpage
\fi

\bibliographystyle{IEEEtran}
\bibliography{mylib}


\end{document}